\documentstyle[prl,aps]{revtex}
\begin{document}
\draft
\title{Tabletop Experiment to Verify Magnetic Monopoles}
\author{Rainer W. K\"uhne}

\maketitle

\begin{abstract}
Recently, we suggested a model of magnetic monopoles (hep-ph/9708394). 
Here we will propose a tabletop experiment to test this model. 
The verification of the predicted effect would have far-reaching consequences.
\end{abstract}

Recently, we suggested a quantum field theoretical model of the 
electromagnetic interaction which includes magnetic monopoles \cite{1}. 
This model is based on Dirac's original proposal of a two-potential 
theory \cite{2} that includes the well-known photon and also Salam's 
``magnetic photon'' \cite{3}. According to this model the cross-section of 
the magnetic photon depends on the absolute motion of the electric charge 
with which it interacts.

Such an absolute rest frame is supported by several observations.

(1) General relativity gives rise to an expanding universe and therefore to a 
finite-sized light zone. The center-of-mass frame of this Hubble sphere can be 
regarded as a preferred frame.

(2) A preferred motion is given at each point of space by cosmological 
observations, namely the redshift-distance relation generated by the 
superposition of the Hubble and the Doppler effect which is isotropic only 
for a unique rest frame. 

(3) A preferred motion is given also by the dipole anisotropy of the cosmic 
microwave background radiation due to the Doppler effect by the Earth's 
motion of 370 km/s \cite{4}.

According to the proposed model this absolute rest frame gives rise to local 
physical effects. The interaction cross-section of a magnetic photon (with 
conventional matter in the terrestrial rest frame) is predicted to be smaller  
than the one of a photon of the same energy, the suppression factor is 
$(370/299,792)^{2}\simeq 10^{-6}$. This means that each reaction that produces 
photons does also create magnetic photons. Furthermore, magnetic photons are 
one million times harder to create, to shield, and to absorb (detect) than 
photons of the same energy.

This model can be tested by a relatively simple experiment. Shoot a laser 
beam on a metal foil with a thickness of several micrometers and place 
a photographic plate (or a charge coupled device or a photomultiplier tube) 
behind the foil to measure the intensity of the penetrating fraction of the 
beam. Our model predicts the detected intensity to be of the order of 
$10^{-12}$ times the one of the original laser beam (only a fraction of the 
magnetic photon part of the beam will be detected, the photon part is 
entirely absorbed by the metal foil). In contrast to this, the standard model 
predicts zero intensity. Background effects by scattering can presumably 
be reduced to several orders below the predicted effect. 

It will be of great benefit to perform the proposed experiment. The 
verification of the predicted effect would have far-reaching consequences. 

(1) Evidence for a new gauge boson and a new kind of radiation (``magnetic 
photon rays'').

(2) Indirect evidence for magnetic monopoles. 

(3) Evidence for an absolute rest frame that gives rise to local physical 
effects.

\end{document}